\documentclass[usenatbib]{mn2e}
\usepackage{longtable}
\usepackage{graphicx}
\usepackage{amsmath}
\usepackage{amssymb}
\usepackage{times}
\usepackage{mathtools}

\usepackage{morefloats}
\usepackage{etex}
\pdfoutput=1

\title[G2C2 I: Homogeneous SDSS photometry for Galactic GCs]{G2C2 I: Homogeneous photometry for Galactic Globular Clusters in SDSS passbands}  
\author[J. Vanderbeke et al.]
{Joachim Vanderbeke$^{1,2}$\thanks{E-mail: Joachimvanderbeke@gmail.com}, Michael
  J. West$^{2,3}$, Roberto De Propris$^4$,  Eric W. Peng$^{5,6}$ , \newauthor
  John P. Blakeslee$^{8,9}$, Andr\'{e}s Jord\'{a}n$^{7}$, Patrick
  C\^{o}t\'{e}$^{8}$, Michael Gregg$^{10,11}$, \newauthor Laura
  Ferrarese$^{8}$, Marianne Takamiya$^{12}$, Maarten Baes$^{1}$ \\ \\
$^{1}$ Sterrenkundig Observatorium, Universiteit Gent, Krijgslaan 281 S9, B-9000 Gent, Belgium\\
 $^{2}$ European Southern Observatory, Alonso de C\'{o}rdova 3107,
 Vitacura, Santiago, Chile\\ 
$^{3}$ Maria Mitchell Observatory, 4 Vestal Street, Nantucket, MA
02554, USA \\
$^{4}$ Finnish Centre for Astronomy with ESO (FINCA), University of Turku, V{\"a}is{\"a}l{\"a}ntie 20,
   FI-21500 Piikki{\"o}, Finland \\
$^{5}$ Department of Astronomy, Peking University, Beijing 100871, China \\
$^{6}$ Kavli Institute for Astronomy and Astrophysics, Beijing 100871,
China \\
$^{7}$ Instituto de Astrof\'isica, Facultad de F\'isica,
Pontificia Universidad
  Cat\'olica de Chile, Av.\ Vicu\~na Mackenna
    4860, 7820436 Macul, Santiago, Chile \\
$^{8}$ Herzberg Insititute of Astrophysics, National Research Council,
Victoria, BC V9E2E7, Canada \\
$^{9}$ Department of Physics and Astronomy,
Washington State University, 1245 Webster Hall, Pullman, WA
99163-2814, USA \\
$^{10}$ Department of Physics, University of California, Davis, CA 956160,
USA\\
$^{11}$ Institute for Geophysics and Planetary Physics, Lawrence
Livermore National Laboratory, L-413, Livermore, CA 94550, USA \\
$^{12}$ Physics and Astronomy Department, University of Hawaii Hilo, Hilo, HI
96720, USA }

\newcommand{\degree}{\ensuremath{^\circ}}

\begin{document}

\date{Accepted. Received }

\maketitle

\label{firstpage}

\begin{abstract}
We present $g^\prime$ and  $z^\prime$ aperture photometry for 96 Galactic Globular Clusters, making 
this the largest homogeneous catalog of photometry for these objects in the SDSS filter system. For a subset 
of 56 clusters we also provide photometry in $r^\prime$ and $i^\prime$. We carry out comparisons with previous 
photometry as well as with the SDSS dataset. The data will be useful for a series of applications in Galactic and 
extragalactic astrophysics. Future papers will analyse the colour-metallicity relation, colour-magnitude diagrams, and structural parameters. The compilation of results based on this dataset will be collected in the Galactic Globular 
Cluster Catalog (G2C2). 
\end{abstract}


\section{Introduction}

Globular clusters (hereafter GCs) formed during the earliest episodes of star formation in galaxies. They are found in all 
but the smallest dwarf galaxies, with massive galaxies hosting systems of hundreds or thousands of clusters. The 
properties of globular clusters appear to be very homogeneous from one galaxy to the other (in terms of colour,
luminosity distribution, etc.) and this implies that the formation of these objects has been intimately related to the
assembly of their parent galaxies (e.g., \citealt{Harris1991}). Globular clusters are living fossils of the Universe at
high redshift (their mass is similar to the Jeans mass at the epoch of recombination) and therefore give a snapshot
of conditions as prevailed at early epochs (see \citealt{West2004,Brodie2006} for reviews). The integrated properties
of globular clusters therefore provide us with information on the earliest stages of galaxy formation; the high intrinsic 
luminosities of clusters means that they can be studied in detail well beyond the Local Group, while the bright end of 
the globular cluster luminosity function has been detected around a $z\sim0.2$ elliptical galaxy \citep{AlamoMartinez2013}.

Most work in both Galactic and extragalactic GCs is still based on the older photometric systems (such as Johnson-Cousins, 
Washington, etc.). Several authors have remarked on the lack of calibrating studies of globular clusters in the SDSS system
\citep{Jordan2005,Sinnott2010,Peacock2011,Vickers2012}; as most such objects are in the South, the vast majority 
of Galactic GCs have not been imaged by the SDSS survey. The latest edition of the \cite{Harris1996} compilation (2010 
edition, this is the version we refer to in the remainder of the paper) lists $UBVRI$ colours for about half of the 150 Galactic 
GCs. Nevertheless, this photometry is inhomogeneous, as it is taken from different papers, using different methods and 
instruments (including photomultipliers, photographic plates and modern CCDs). 

The Sloan Digital Sky Survey (SDSS -- \citealt{York2000}) has now imaged over a quarter of the Northern sky (about 14500 
square degrees) in five passbands. Together with upcoming imaging surveys in the South, SDSS will completely replace the 
older Schmidt plate atlases of the sky, and at the same time provide a standardized system of photometry in the optical for 
astrophysics (theoretically, with calibrators in every field). With this motivation, our team embarked on the Galactic Globular 
Cluster Catalog (G2C2) project, with an ultimate goal of collecting reliable photometry using the SDSS filter system for a 
large sample of Galactic GCs.  In this first paper, we present $g^\prime$ and $z^\prime$ magnitudes for about two-thirds 
of the Galactic GCs and $r^\prime$ and $i^\prime$ magnitudes for about one-third of all Galactic GCs. Future work will discuss the colour-metallicity relation (see the companion Paper II -- Vanderbeke et al. 2013, in press), the 
colour-magnitude diagrams of these clusters, their spectral energy distributions over 2 decades in wavelength, and the 
structural parameters of GCs using King models. 

Here we discuss the buildup of the photometric database: imaging of 96 Galactic clusters in at least 2 SDSS bands (as well
as 2 more for a subset of 56 objects). We describe our observations and basic data reduction: we give details about the
samples, determination of cluster centres, aperture photometry, estimation of the sky level, removal of outliers, photometric
errors and correction for extinction. To assess the quality of our data we compare these
with previous work and carry out a similar analysis on globular clusters in common with the SDSS footprint. This paper is 
organized as follows. Section~\ref{sec:data} presents the observations and the basic data reduction. We present the 
integrated photometry and colours for the Galactic GCs in Section~\ref{sec:intphotGC}.  We summarize the results in Section~\ref{sec:results}.

\section{Observations and data reduction}
\label{sec:data} 
\subsection{CTIO}
We selected Galactic GCs from the latest versions of the \cite{Harris1996} catalog, which includes about 150 GCs. 
Observations were carried out between 2003 May 10 and 2012 June 9 using the CTIO 0.9~m and 1~m telescopes with 
the USNO $g^\prime r^\prime i^\prime z^\prime$ filter set. Because the Galactic bulge and hence the bulk of the Galactic 
GCs are best observable during the Chilean winter, cirrus and bad weather were a real issue during the observing runs: 
many nights were totally lost due to clouds or strong winds, while other nights were disturbed by cirrus and were not
photometric. 

For the results in this paper, we reduced 13 nights of observations. Several clusters were observed multiple times and it 
became clear that only 4 nights (all of which used the 0.9 m telescope, with an instrumental set-up as shown in 
Table~\ref{tab:setup}) could be considered (largely) photometric. During these nights, we collected $g^\prime$ and 
$z^\prime$ observations for 81 GCs, about half of which we also observed using $r^\prime$ and $i^\prime$ filters.

\begin{table}
\centering
\caption{\label{tab:setup} The instrumental set-up. }
\begin{tabular}{ll}
\hline
Telescope & CTIO 0.9m \\
Dates & 2003 May - 2012 March  \\
\hline
Filter set & $g^\prime r^\prime i^\prime z^\prime $ \\
Spatial scale & 0.396 \arcsec~pixel$^{-1}$\\
Field Size & 13.6\arcmin$\times$13.6\arcmin \\
Gain & 3.0 e$^{-}$~ADU$^{-1}$\\
Read-out noise & 5 e$^{-}$\\
Detector & 2048$\times$2046 Tek2K CCD\\
\hline 
\end{tabular} \\
\end{table}

For the vast majority of the clusters, we have 60~s exposures in $g^\prime$ and $z^\prime$ taken in June 2004. During 
the run performed on 2003 May 10 short (between 5 and 30~s) and long (270~s in $g^\prime$, 410~s in $z^\prime$)
exposures were obtained. Both shorter and longer exposures were used separately to determine magnitudes. Some of 
the clusters have very bright stars close to their centres. These very bright stars saturated the CCD even for the short 
exposures. For these clusters (NGC~6397, 47~Tuc, NGC~6121) we obtained additional 1 s exposures. The 
observations discussed in this paper were performed between 2003, May 10 and 2005, September 26 under seeing conditions varying between 1\arcsec~ and $\sim2^{\prime\prime}$.

The basic data reduction was performed via a dedicated \textsc{idl} pipeline developed by our team. The procedure
largely follows conventional CCD reduction processes. The bias level was estimated separately for each quadrant of
the CCD, by computing the median of the corresponding bias section, which was then subtracted for each quadrant. 
The frames were then flat-fielded by the median of the twilight flats taken each night and corrected for bad columns. 
To identify and robustly remove cosmic rays, we used the L.A. Cosmic (imaging version) method \citep{vanDokkum2001}. 

One additional complication was the incorrect information in the fits headers of the clusters observed from 2005 onwards.
John Subasavage (private communication) confirmed that, since the TCS upgrade in early 2005, the header values (including 
RA, DEC, airmass and epoch) are not correct. Based on the coordinates obtained from \cite{Harris1996} 
and the header values (date and time of observation), we computed automatically the airmasses for the observations taken 
after 2005. Comparison with observation log sheets showed excellent agreement.

During the course of each observing night a minimum of several dozen standard stars, selected from \cite{Smith2002}, were observed 
at different airmasses. Photometric calibration (i.e., determination of zeropoints, colour terms and 
atmospheric extinction values, as well as removal of other instrumental signatures) was carried out as in \cite{Patat2001}. 
Foreground (galactic) extinction was estimated for each position using the most recent values from the recalibration of
\cite{Schlafly2011}. 

\subsection{SDSS Data Release 9}

We also considered a sample of 21 northern hemisphere clusters from the ninth data release of SDSS \citep{Ahn2012}, 6 of
which have also been observed by us with the CTIO 0.9m telescope. However, NGC~6838 and NGC~6254 were only partly 
covered by SDSS and were not included in this study, as our procedure (see below) requires us to cover at least the 
half-light radius in each object. Bright foreground stars outshine GLIMPSE01, Ko~1 and Ko~2. We do not consider these 
clusters further.

For some other clusters, several SDSS stripes needed to be assembled into mosaics using
Montage\footnote{http://montage.ipac.caltech.edu/}, although this may lead to issues with variable sky levels. Although
SDSS data have the considerable advantage of being photometrically homogeneous and uniform, the 53.9~s standard exposure
in SDSS saturates bright red giant branch (RGB) stars in some GCs, an effect which becomes clear when comparing the colour
magnitude diagrams and which is further discussed in Section \ref{sec:CTIOSDSS}. Note that the 'SDSS' filters at the APO
2.5m telescope (and the CTIO 0.9m) have significantly different effective central wavelengths from the calibrating filters at 
the USNO 1m telescope, where the $u^\prime g^\prime r^\prime i^\prime z^\prime$ photometric system was defined \citep{Fukugita1996} and extended with secondary standards by \cite{Smith2000,Smith2002}. The conversion between the
$u^\prime g^\prime r^\prime i^\prime z^\prime$ and $ugriz$ system is given on the SDSS
website\footnote{http://www.sdss.org/dr6/algorithms/jeg\_photometric\_eq\_dr1.html}. These result in negligible changes 
to the GC colours when compared to the magnitude uncertainties.

For NGC~6341 and NGC~5904 the saturation of the SDSS chip was so 
severe that almost the entire RGB is brighter than the saturation limit of the CCD. It was nearly impossible to select 
non-saturated stars to construct the PSF for the CMDs, which are indispensable in the reduction process (see Section~\ref{sec:removefgstars}). We decided to discard these clusters from the sample. Nevertheless,
good quality data for NGC~6341 is highly desirable, as this GC is one of the most metal-poor GCs of the Milky Way.

\section{Integrated Photometry of Globular Clusters} \label{sec:intphotGC}

\begin{figure}
\centering 
\includegraphics[scale=0.4,trim=0cm 6.7cm 0cm 6cm] {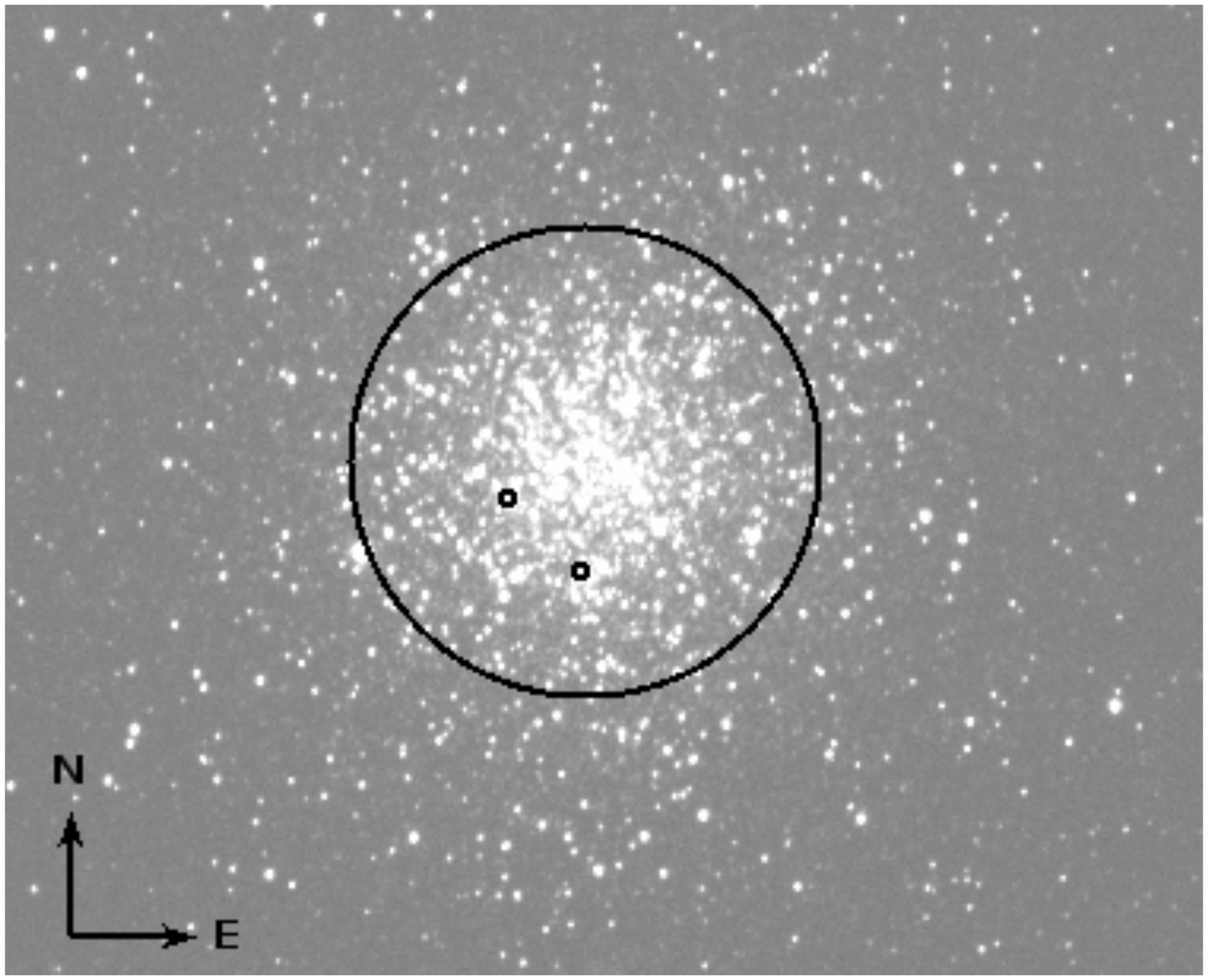}
\caption{$g^\prime$-band observation of NGC~5986. The half-light radius is
  indicated with the big circle, while the small circles indicate two stars that are likely
non-members (contaminants) as selected from their position in the cluster colour-magnitude diagram.}
\label{CTIO_NGC5986}
\end{figure}

Galactic GCs are generally too large on the sky to be completely included within a single CCD frame (see
Figure~\ref{CTIO_NGC5986} for an example from our own data).  
Although we experimented with fitting King models to the surface 
brightness profiles of the GCs to measure total magnitudes (we discuss this in a subsequent paper), we ultimately chose to
derive aperture magnitudes within the clusters' half-light radii (e.g., \citealt{Peng2006}) to determine integrated colours.
As long as the clusters
show no strong colour gradients in their outskirts, the integrated colours we present here should be suitable proxies for studies
of extragalactic systems as well.

We measured the aperture magnitude within the half-light radius $r_h$ obtained from \cite{Harris1996} where the original values are drawn largely from \cite{Trager1993, Trager1995} and \cite{McLaughlin2005}. As the largest $r_h$ is 5 arcminutes, it does
not completely fill the CCD field-of-view and therefore enables us to determine magnitudes for all clusters in a homogeneous 
manner (as in \citealt{Peng2006}). To obtain a total magnitude for the clusters, we would need imaging reaching well beyond the 
tidal radius. However, the largest tidal radius for the Galactic GCs is 53.8 arcminutes, much bigger than our field-of-view. As
long as colour gradients in the cluster outskirts are not very strong, the integrated colours determined within the 1/2 light
radius aperture should be representative of the total colours. To illustrate this for a cluster with a tidal radius $r_t$ which 
fits the CTIO 0.9m telescope field-of-view, we compared the $g^\prime-z^\prime$ colour based on $r_h$ and $r_t$ apertures 
for NGC~5694. After correcting for contaminants (as will be described in Section~\ref{sec:removefgstars}) the colour difference 
between $r_h$ and $r_t$ apertures is 0.002 mag for this cluster, which is negligible compared to the magnitude uncertainties. 

NGC~6287 and NGC~6553 are other clusters with a tidal radius small enough to be entirely covered by the CTIO field-of-view. However, NGC~6287 was not centred properly on the chip and was not entirely covered as a consequence . The observations of NGC~6553 included some saturated stars within the tidal radius, which is complicating the cleaning of the contaminants and impeding a proper comparison of the colours within the half-light and tidal radii.

Our first step will be to determine the cluster centres for the apertures, followed by estimation of the sky values, removal of
contaminants (foreground stars) and measurement of the total flux within the 1/2 light radius. We then discuss extinction,
photometric errors and compare our results with previous work and SDSS.

\subsection{Cluster Centroids}
Our first step is to determine an accurate cluster centroid for the apertures.  For the CTIO data, we followed the method 
of \cite{Bellazzini2007}. We calculated the aperture fluxes in a grid of 25 points around the initial (visual) guess for the 
cluster centre: the size of the grid is 125 pixels or $60''$. The 'centre' position at which the aperture flux is maximal is 
adopted as the cluster centroid and used as the reference point for  aperture photometry. This was performed separately 
for all frames, because of inconsistent coordinates in the headers. For the SDSS data, the coordinates listed in 
\cite{Harris1996} were adopted together with the SDSS astrometric solution. The error in the photometry introduced by 
the uncertainty in the centre position is estimated in the following way (both for SDSS and CTIO data): magnitudes 
were computed for apertures centered at four grid points separated by 5\% of the stated half-light radius; we calculated 
the magnitude difference between the aperture magnitudes centered on these positions and the 'true' centre we determined 
above. The median of this difference is used as the estimate of the uncertainty introduced by the centre determination and 
is summed in quadrature to the photometric and other errors to obtain the  total magnitude error, assuming these errors 
are independent. The median contribution of the centre determination to the total magnitude uncertainty is 
0.006 mag.

To provide a consistency check, we have compared the centroids in the $g^\prime$-band to the more accurate
central coordinates for each cluster as determined by \cite{Goldsbury2010} with HST/ACS data. The median difference 
between both centre determinations is $0.086~r_h$. Using $8.6\% r_h$ instead of $5\% r_h$ to determine the magnitude
uncertainty due to the centre determination results in a median additional error of 0.004 mag, which is negligible when 
compared to the systematic error introduced in Section~\ref{sec:diffnights}.

\subsection{Sky values} \label{sec:sky}
Determination of the sky value proved challenging, as several clusters fill the 0.9m CCD and in most cases the images 
do not cover the clusters out to their tidal radius. We used \textsc{mmm} (Mean, Median, Mode), a routine available 
at the \textsc{idl} astronomy library which was developed to estimate the sky background in a crowded field and was 
adapted from the \textsc{daophot} routine with the same name. The algorithm consists of several steps: it 
first computes the mean and standard deviation of the sky flux, which is used to eliminate outliers. \textsc{mmm} repeats 
the first step in up to 30 iterations recomputing the sky (eliminating outliers of the previous iteration). As a next step, \textsc{mmm} estimates the amount of  stellar contamination by comparing the mean, mode and median of the remaining 
sky pixels. If the mean is less than the mode and the median, then the contamination is slight and the sky is estimated 
by the mean. If the mean is larger, indicating severe contamination (as the program assumes positive departures from the 
true sky value in crowded fields), then the true sky value is estimated by $3\times \text{median}-2\times\text{mean}$. 
We applied the \textsc{mmm} method to the four corners of each frame in a 100 by 100 pixel area, summing up about 
40000 pixels (as some pixels will be identified as outliers by \textsc{mmm} and will not contribute to the sky determination). 
We regard these regions are the best approximation for the sky value.

The SDSS pipeline processing the data includes the sky subtraction, hence the sky value for these frames is always about zero. For consistency, we did determine the sky value running \textsc{mmm} on the entire mosaic. 

Photometric uncertainties introduced when determining the sky level are further discussed in Section~\ref{sec:diffnights}. 

\subsection{Removing foreground stars} \label{sec:removefgstars}

Contamination from foreground bulge or disk stars can be severe at low Galactic latitudes and a number of methods have been proposed to 
tackle this issue. In their study on the integrated 2MASS photometry of Galactic GCs, \cite{Cohen2007} considered stars 
brighter than the tip of the RGB by 1.5 mag as non-members and excluded them. However, it is difficult to use this 
approach close to the cluster centre (at least from the ground) because of crowding and the low spatial resolution of their 
(and our) data. \cite{Peng2006}, for example, disregarded this correction. 

Contaminating stars can be excluded in two ways: from their abnormal position in the cluster colour-magnitude diagram, 
which implies they are unlikely to be cluster members (e.g., if they lie well outside the cluster principal sequences), or from 
their measured proper motions, as cluster stars are unlikely to show detectable motions because of their great distances. 

Although our data suffer from crowding, average seeing and poor spatial resolution, we were able to derive colour-magnitude
diagrams to identify likely foreground stars and clean the aperture magnitudes. We carried out stellar photometry with
\textsc{daophot} and \textsc{allstar} \citep{Stetson1987,Stetson1994}. As a first step, up to 50 isolated and bright stars 
were selected to model a point spread function (PSF), accounting for variation over the field by allowing quadratic variability. 
We used a PSF radius depending on the seeing: generally we  used $4\times$FWHM but adopted a maximum of 15 pixels 
when the seeing was bad or the focus was mediocre. This aperture is large enough to remove the bulk of the contaminating 
star light but small enough to enable \textsc{daophot} to resolve the stars. \textsc{daomatch} and \textsc{daomaster} were 
used to crossmatch the different filters.

As a consequence of the low resolution of our data (0.396\arcsec pixels and a seeing between 1\arcsec~ and $\sim2^{\prime\prime}$), crowding does obviously affect the final CMDs, which are not complete, especially close to the centre. However, these cover large 
fields, extending well beyond the half-light radius where crowding is not as important. A detailed analysis of these colour-magnitude
diagrams will be presented in a forthcoming paper. 

\begin{figure}
\includegraphics[scale=0.87,trim=2.8cm 13.1cm 9cm 6cm] {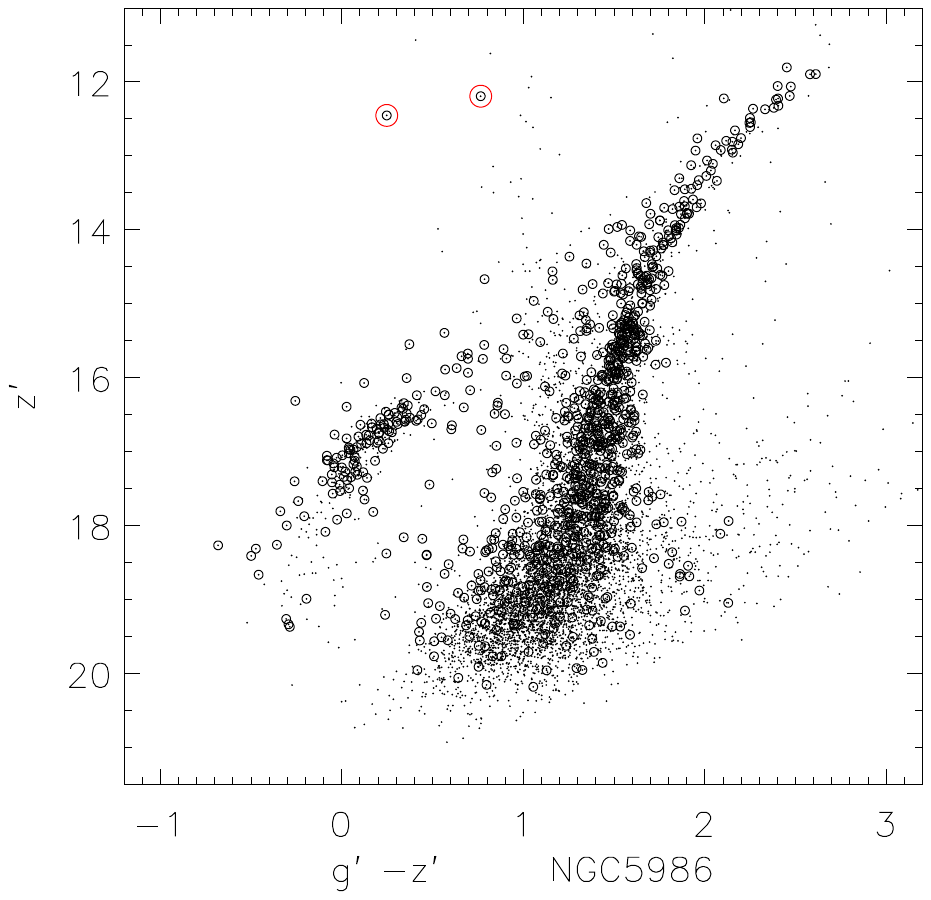}
\caption{NGC~5986: open circles are stars within $r_h$, dots are stars from the
  entire field. The two bright isolated stars in the CMD (indicated
  with red circles) are selected
  as candidate outliers and were checked for proper motions. See text
  for more details.   }
\label{CMD_NGC5986}
\end{figure}

As an example, Fig.~\ref{CMD_NGC5986} shows the CMD for NGC~5986: open circles represent stars within the 
half-light radius, dots are stars from the entire field. This GC is located at a Galactic latitude of $b=13.27\degree$, so 
some contamination from the  disk may be expected, and is visible as a blue plume of stars above the turnoff. As mentioned 
above we use the half-light radius to measure the aperture magnitudes of the cluster, so stars in this area (represented as
open circles) that lie outside of the principal sequences are possible contaminants. For confirmation, we checked 
the bright outliers, indicated by red circles in Fig.~\ref{CMD_NGC5986}, for proper motions in the USNO-B1.0 catalog
\citep{Monet2003} and the NOMAD catalog \citep{Zacharias2005}, although in many cases these are not fully conclusive 
(e.g., see \citealt{McDonald2013} for a similar approach to the bright AGB stars in NGC 4372). Once we are convinced that 
the star is a true non-member, the star is cleaned from the cluster photometry by subtracting its flux, based on the
\textsc{daophot} PSF magnitudes, from the flux in the cluster aperture.  Removing these stars in NGC~5986 results in
magnitude corrections of 0.06 (0.03, 0.02, 0.02) in $g^\prime$ ($r^\prime$, $i^\prime$, $z^\prime$, respectively). It is
interesting to note that the contaminating stars in NGC~5986 would not have been removed if we had followed 
\cite{Cohen2007} as they are fainter than the RGB tip.

The magnitude corrections for foreground contamination may sometimes be very large, especially in poor clusters: for 
Pal~10 these corrections are 1.33 (0.62, 0.39, 0.33) mag. in $g^\prime$ ($r^\prime$, $i^\prime$, $z^\prime$). This 
yields a $\sim1$ mag correction for contamination in $g^\prime-z^\prime$. 

A caveat is that differential reddening may shift foreground stars into the cluster principal sequences: this can be
significant for clusters at low galactic latitude, where extinction may be patchy \citep{AlonsoGarcia2012}. While we
discuss reddening related issues extensively in our study on the colour-metallicity relation, specifically as these affect the colour-magnitude relation, we believe 
that a few such outliers will not significantly affect the derived colours.

For the CTIO data, an extract of the magnitudes and the applied contamination corrections (denoted as CMD$_{g^\prime,r^\prime,i^\prime,z^\prime}$) is listed in Table~\ref{tab:griz}. The complete table is available in the electronic version of this paper. Magnitudes and contamination corrections for the SDSS data are given in Table~\ref{tab:SDSSdr9}. 

\begin{table*}
\centering
\caption{\label{tab:griz} Extract of the GC $g^\prime r^\prime i^\prime z^\prime$ magnitudes and errors based on CTIO observations.  CMD$_{g^\prime,r^\prime,i^\prime,z^\prime}$ presents the magnitude corrections based on the CMDs. The complete table is available in the online version of the paper. }
\begin{tabular}{lcccccccccccc}
\hline
ID & $g^\prime$ & $\sigma_{g^\prime}$ & $r^\prime$ & $\sigma_{r^\prime}$ & $i^\prime$ & $\sigma_{i^\prime}$ & $z^\prime$ & $\sigma_{z^\prime}$ &CMD$_{g^\prime}$ & CMD$_{r^\prime}$ & CMD$_{i^\prime}$ & CMD$_{z^\prime}$ \\
\hline
NGC104     &      4.912     &      0.030     & ...     & ...     & ...     & ...     &      3.677     &      0.044     &      0.00     & ...     & ...     &      0.00 \\
 NGC288     &      9.080     &      0.032     &      8.600     &      0.045     &      8.295     &      0.044     &      8.139     &      0.044     &      0.00     &      0.00     &      0.00     &      0.00 \\
 NGC362     &      7.471     &      0.030     &      6.925     &      0.043     &      6.618     &      0.043     &      6.419     &      0.043     &      0.01     &      0.02     &      0.02     &      0.03 \\
NGC1261     &      9.474     &      0.031     &      8.995     &      0.045     &      8.744     &      0.045     &      8.582     &      0.045     &      0.03     &      0.05     &      0.06     &      0.06 \\
    AM1     &     15.958     &      0.035     &     15.582     &      0.046     &     15.219     &      0.047     &     15.159     &      0.056     &      0.07     &      0.13     &      0.14     &      0.16 \\
NGC1851     &      8.280     &      0.031     &      7.703     &      0.044     &      7.413     &      0.044     &      7.194     &      0.044     &      0.03     &      0.04     &      0.05     &      0.05 \\
NGC1904     &      9.006     &      0.030     &      8.570     &      0.044     &      8.351     &      0.044     &      8.178     &      0.044     &      0.02     &      0.04     &      0.05     &      0.05 \\
NGC2298     &      9.650     &      0.039     &      9.287     &      0.050     &      9.036     &      0.052     &      8.865     &      0.054     &      0.06     &      0.10     &      0.10     &      0.10 \\
NGC2808     &      6.695     &      0.030     &      6.194     &      0.044     &      5.916     &      0.044     &      5.725     &      0.044     &      0.00     &      0.00     &      0.00     &      0.00 \\

\hline 
\end{tabular} \\
\end{table*}

\begin{table*}
\caption{\label{tab:SDSSdr9} GC $griz$ magnitudes and errors based on data from SDSS (Data Release 9).  CMD$_{g,r,i,z}$ presents the magnitude corrections based on the CMDs. }
\centering
\begin{tabular}{lccccccccccccc}
\hline
ID &  $g$ & $\sigma_g $& $r$ & $\sigma_r$ & $i$ &$\sigma_i$ &$z$ & $\sigma_z$ & CMD$_g$& CMD$_r$& CMD$_i$& CMD$_z$\\ \hline
Whiting1 &          16.637 &           0.083 &          16.162 &           0.069 &          15.953 &           0.060 &          15.854 &           0.060 &            0.00 &            0.00 &            0.00 &            0.00 \\ 
Pal1 &          15.118 &           0.042 &          14.769 &           0.033 &          14.615 &           0.030 &          14.522 &           0.032 &            0.46 &            0.62 &            0.67 &            0.69 \\ 
NGC2419 &          11.198 &           0.009 &          10.727 &           0.009 &          10.473 &           0.009 &          10.371 &           0.010 &            0.01 &            0.01 &            0.01 &            0.01 \\ 
Pal3 &          15.389 &           0.046 &          14.824 &           0.036 &          14.585 &           0.033 &          14.304 &           0.043 &            0.01 &            0.03 &            0.04 &            0.04 \\ 
Pal4 &          15.284 &           0.052 &          14.636 &           0.058 &          14.293 &           0.083 &          14.074 &           0.107 &            0.00 &            0.00 &            0.00 &            0.00 \\ 
NGC4147 &          11.185 &           0.010 &          10.823 &           0.008 &          10.670 &           0.007 &          10.443 &           0.005 &            0.00 &            0.00 &            0.00 &            0.00 \\ 
NGC5024 &           8.518 &           0.003 &           8.122 &           0.004 &           7.991 &           0.003 &           7.676 &           0.003 &            0.00 &            0.00 &            0.00 &            0.00 \\ 
NGC5053 &          10.734 &           0.012 &          10.379 &           0.008 &          10.164 &           0.007 &          10.062 &           0.005 &            0.00 &            0.00 &            0.00 &            0.00 \\ 
NGC5272 &           7.071 &           0.002 &           6.712 &           0.002 &           6.555 &           0.001 &           6.135 &           0.002 &            0.00 &            0.00 &            0.00 &            0.00 \\ 
NGC5466 &          10.180 &           0.013 &           9.725 &           0.011 &           9.466 &           0.009 &           9.413 &           0.009 &            0.00 &            0.00 &            0.00 &            0.00 \\ 
Pal5 &          12.383 &           0.069 &          11.941 &           0.067 &          11.800 &           0.056 &          11.483 &           0.136 &            0.07 &            0.10 &            0.14 &            0.14 \\ 
Pal14 &          14.654 &           0.028 &          13.937 &           0.047 &          13.689 &           0.044 &          13.641 &           0.052 &            1.02 &            0.58 &            0.79 &            0.81 \\ 
NGC6205 &           6.903 &           0.005 &           6.667 &           0.003 &           6.397 &           0.004 &           6.076 &           0.002 &            0.00 &            0.00 &            0.00 &            0.00 \\ 
NGC6229 &          10.419 &           0.007 &           9.876 &           0.005 &           9.751 &           0.005 &           9.489 &           0.004 &            0.00 &            0.00 &            0.00 &            0.00 \\ 
Pal15 &          13.618 &           0.032 &          13.072 &           0.043 &          12.769 &           0.060 &          12.527 &           0.057 &            0.19 &            0.16 &            0.14 &            0.12 \\ 
NGC6535 &          10.274 &           0.030 &           9.901 &           0.029 &           9.773 &           0.033 &           9.470 &           0.040 &            0.11 &            0.08 &            0.06 &            0.06 \\ 
NGC6934 &           9.446 &           0.003 &           9.059 &           0.003 &           8.858 &           0.003 &           8.576 &           0.003 &            0.00 &            0.00 &            0.00 &            0.00 \\ 
NGC7006 &          11.320 &           0.012 &          10.817 &           0.012 &          10.554 &           0.012 &          10.436 &           0.012 &            0.02 &            0.02 &            0.02 &            0.02 \\ 
NGC7078 &           7.230 &           0.006 &           6.876 &           0.004 &           6.802 &           0.004 &           6.289 &           0.008 &            0.00 &            0.00 &            0.00 &            0.00 \\ 
NGC7089 &           7.298 &           0.003 &           6.854 &           0.004 &           6.758 &           0.004 &           6.407 &           0.002 &            0.00 &            0.00 &            0.00 &            0.00 \\ 
Pal13 &          15.736 &           0.060 &          15.392 &           0.050 &          15.270 &           0.048 &          15.147 &           0.049 &            0.00 &            0.00 &            0.00 &            0.00 \\ 
\hline 
\end{tabular} \\
\end{table*}

\subsection{Comparison with previous work}
\begin{figure*}
\centering 
\includegraphics[scale=0.95,trim= 3cm 13.7cm 0 6cm] {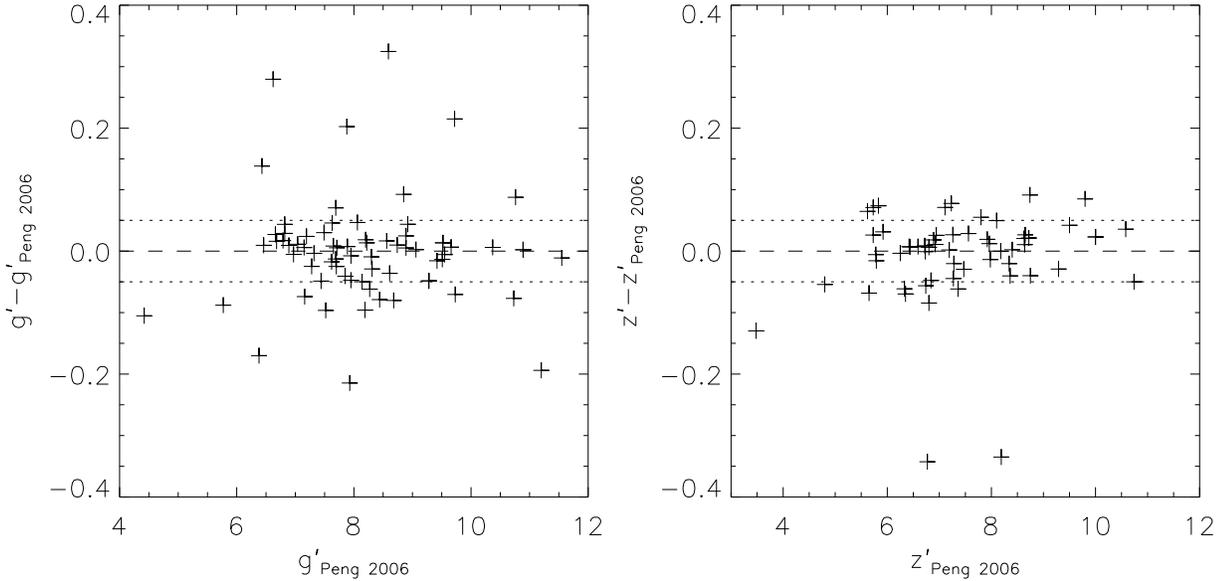}
\caption{Comparison of the $g^\prime$ and $z^\prime$ magnitudes with the magnitudes obtained by Peng et al. (2006).}
\label{comparewithPeng}
\end{figure*}

To test the reliability of our approach we compare our $g^\prime$ and $z^\prime$ magnitudes with \cite{Peng2006}, which also uses the half-light radius. Moreover, their Galactic GC data was based on the observations performed on 2004 June 5 and 6, so there is a considerable overlap with our sample. We match our apertures to theirs (some measurements of structural parameters have since
changed) and we use the $E(B-V)$ values from \cite{Harris1996} which were used by \cite{Peng2006} and a \cite{Cardelli1989} 
reddening law instead of the reddening values from \cite{Schlafly2011}, for the sake of consistency in these comparisons. These 
are shown in Fig.~\ref{comparewithPeng}: while there is no systematic offset, there are a couple of outliers, for which the magnitude
difference with the earlier results \citep{Peng2006} is larger than expected. It is unclear what the origin of the discrepancy
is.  In the next section, we discuss the origin of the photometric errors in more detail. It will become clear that the sky determination can strongly affect the final magnitudes, which we raise as possible cause for the variance when comparing to \cite{Peng2006}.
At least for NGC5927 the observing log of the original \cite{Peng2006} hints at clouds or cirrus and this may be another possible
reason for the difference. The RMS scatter of our photometry, compared with \cite{Peng2006}, is 0.08 (0.07, 0.09) for $g^\prime$ ($z^\prime, g^\prime-z^\prime$).

\subsection{Photometric errors} \label{sec:diffnights}

\begin{figure*}
\includegraphics[scale=0.8,trim= 1.5cm 13.4cm 4 6cm] {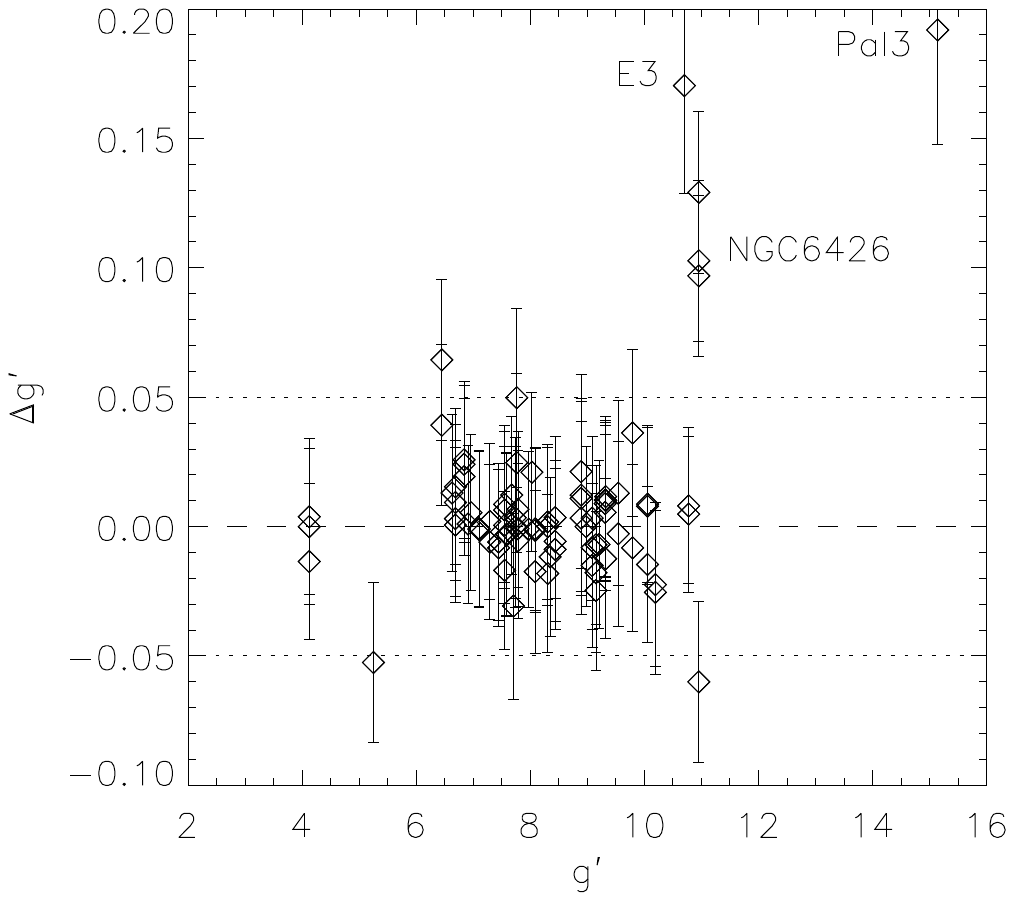}\includegraphics[scale=0.8,trim= 11.3cm 13.4cm 4 6cm]
{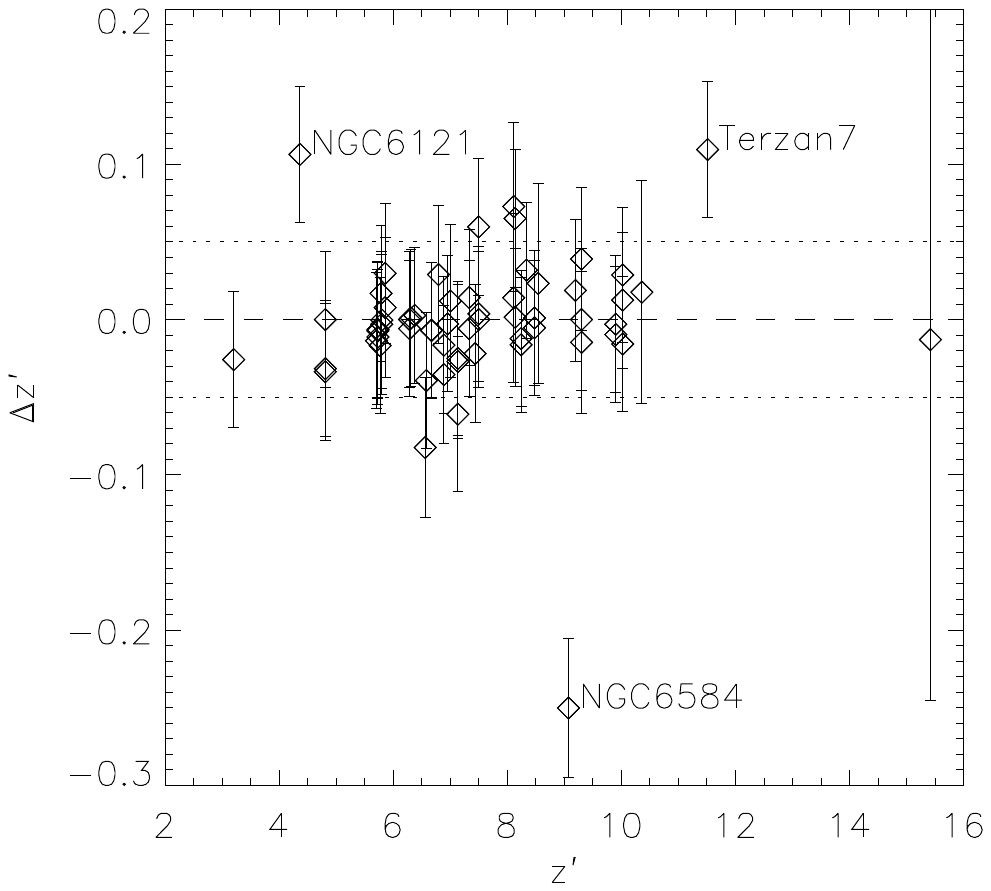}
\caption{Comparison of $g^\prime$ and $z^\prime$ magnitudes, based on observations performed on different nights or performed subsequently during the same night). Error bars
  are including the systematic error. }
\label{compdiffnight}
\end{figure*}

Except for some poor clusters, the integrated magnitudes over the half-light radius have small random errors. The main
contributions to the photometric error budget  come from uncertainties in the photometric calibration and the centre
determination (discussed above). For several clusters we have $g^\prime$ and $z^\prime$ data available from consecutive short and long exposures or from observations obtained during different 
nights. In this case the median magnitude of all observations is taken as the final magnitude in Table~\ref{tab:griz}. In
Fig.~\ref{compdiffnight} we compare the magnitude differences between different observations (both performed on different nights or subsequent observations performed during the same night). The magnitudes in general 
compare well. However, there are some exceptions which will be treated later in this section. 

First, we introduce some parameters that will be used to describe the details of these clusters. The standard way adopted in this study to estimate the sky level was described in Section~\ref{sec:sky}. We now refer to this sky determination as method A. To test the influence of the sky determination on the cluster magnitude, we estimate the sky contribution using a sky band, centred in the cluster centre, with an inner radius of 900 pixels and a width of 100 pixels. We refer to the latter as method B. The resulting magnitude difference between method A and method B is denoted as $\Delta_{AB} ^{*}$ (with '*' indicating the filter). Tests on SDSS data did not result in significantly different sky estimates using methods A and B. 

Another parameter used in the remainder of this section is $R_{GC/Sky,*}$, which is the ratio of the sky-subtracted cluster flux to the sky flux (both measured within a half-light radius). Hence, when $R_{GC/Sky,*}=1$, the sky contribution is as strong as the cluster contribution to the flux within $r_h$.  $R_{GC/Sky,*}<1$ when the sky contribution is higher than the pure (sky-subtracted) cluster contribution.

E~3 is a faint old cluster (12.8 Gyr, \citealt{MarinFranch2009}) that may have been truncated by tidal forces \citep{vandenbergh1980}. Observations of 60~s were performed on both June 5 2004 and September 26 2005, with magnitude differences between both nights of 0.17 mag in $g^\prime$ and 0.02 mag in $z^\prime$. It is suspicious that the magnitudes compare well for the $z^\prime$ filter, while they do not for the $g^\prime$ filter. In the latter filter, the average cluster surface brightness within a half-light radius is much lower than the brightness of the sky. Moreover, for the observations performed on June 5 2004, $R_{GC/Sky,g}= 0.18$, while on September 26 2005 $R_{GC/Sky,g}= 0.07$. This is a first indication the magnitude difference could be attributable to the sky uncertainty. Using method B (with the sky radius) instead of method A (with the four corners) results in an absolute $g^\prime$ magnitude difference $|\Delta_{AB} ^{g^\prime}|\sim0.1$ mag, while the magnitude uncertainty is about 0.01 mag. This shows that the contribution of the sky uncertainty to the magnitude uncertainty is underestimated. Remark that the magnitude resulting using method B instead of method A was 0.1 mag brighter for the June 5 2004 observation ($\Delta_{AB} ^{g^\prime}\sim0.1$ mag), while it was 0.1 mag fainter for the September 26 2005 observation ($\Delta_{AB} ^{g^\prime}\sim-0.1$ mag). For the $z^\prime$ filter,  $R_{GC/Sky,z}$ is about 0.02 for both nights. The $z^\prime$ magnitudes are differing for both nights by $\Delta_{AB} ^{z^\prime}\sim-0.1$ mag when using a sky band instead of the 4 corner approach. \cite{Peng2006} did not include E~3 in their analysis because of the significant reddening ($E(B-V)\sim0.3$, \citealt{Harris1996}). 

NGC6426, a high-reddening cluster with $E(B-V)\sim0.36$, was observed on May 10, 2003 and on June 5 and 6, 2004. The $g^\prime$-band magnitudes show significant variations between the nights: magnitudes obtained on May 10, 2003 were about 0.1 mag brighter than for the observations performed in June 2004. Taking the median of the observations performed on the different nights resulted in a $g^\prime$ magnitude consistent with all the June 2004 observations. The resulting $z^\prime$ magnitudes are all within 0.04 mag. Again, we want to pinpoint the influence of the sky determination on the magnitude discrepancies. The ratio $R_{GC/Sky,g}$ is about 0.6 for the observations performed on May 10 2003, while it is about 0.05 for the June 5 and 6 2004 observations. $\Delta_{AB} ^{g^\prime}\lesssim 0.02$ mag for the May 10 2003 observations, while these differences are as high as 0.34 mag (0.24 mag) for the June 5 (6, respectively) 2004 observations. For the 75~s and 410~s $z^\prime$ observations taken on May 10 2003, the ratio $R_{GC/Sky,z}\sim0.1$, with $\Delta_{AB} ^{z^\prime}< 0.03$. For the 60~s observation performed on June 5 2004, $R_{GC/Sky,z}\sim0.14$ and $\Delta_{AB} ^{z^\prime}\sim0.24$, while for the 410~s observation taken on June 6 2004, $R_{GC/Sky,z}\sim0.08$ and $\Delta_{AB} ^{z^\prime}\sim0.15$. However, pure Poisson magnitude errors are smaller than 0.01. It is clear that some of the variation over the different nights can be attributed to the sky determination for this cluster.    

Pal~3 is a low-reddening cluster ($E(B-V)\sim0.04$) and was observed on June 5 and 6 2004. The magnitude difference in $g^\prime$ was about 0.19 mag (and 0.01 mag in $z^\prime$). A bright star on the edge of the aperture complicates the determination of the magnitude. $R_{GC/Sky,g}$ amounts about 0.05 on both nights, while $R_{GC/Sky,z}$ varies between 0.002 and 0.02, so again the cluster is much fainter than the sky level. Using a sky band with \textsc{mmm} to estimate the sky value, results in $g^\prime$ ($z^\prime$) magnitude differences up to $\Delta_{AB} ^{g^\prime}\sim1$ mag ($\Delta_{AB} ^{z^\prime}\sim3.4$ mag, respectively). The sky values obtained with \textsc{mmm} in the corners (as described in Section~\ref{sec:sky}) and the sky band are consistent within the large uncertainties. However, the photometric magnitude uncertainties given by the routine are large (up to $\sigma_{z^\prime}\sim0.23$), but not as large as the differences invoked by different sky estimation methods.

NGC~6584 was observed on May 10 2003 and June 5 2004. While $g^\prime$ magnitudes compare well for both nights, $z^\prime$ magnitudes show larger differences. For the $z^\prime$ observations performed on June 5, $R_{GC/Sky,z}$ is 0.86, while for the observations taken on May 10 it amounts to 0.60. $\Delta_{AB} ^{z^\prime}\sim0.04$ mag for the June 2004 observation, while $\Delta_{AB} ^{z^\prime}\sim-0.02$ mag for the May 2003 observation. For this cluster, the sky determination seems not to be responsible for the magnitude offset. As a reference, we give some details on the $g^\prime$ observations as well. For the 37~s and 270~s observations performed on May 10 2003, $R_{GC/Sky,g}\sim6$ and $\Delta_{AB} ^{g^\prime}<0.01$ mag. For the observation obtained on June 5 2004, $R_{GC/Sky,g}\sim0.53$ and the magnitude difference (using a sky band instead of the corners) is about $\Delta_{AB} ^{g^\prime}\sim$ 0.03 mag.

Another cluster with a large difference between different observations is Terzan~7. This faint cluster was only observed on May 10 2003, but consecutive $z^\prime$-band 79~s and 410~s observations showed magnitude variations about 0.1 mag. Because it is unlikely that, during a night that is considered photometric, the observing conditions change drastically in a ten minute timespan, this magnitude difference is surprising and deserves some special attention. This cluster, at a Galactic latitude of $b\sim-20\degree$, is associated with the Sagittarius stream, hence located in a crowded field. This obviously complicates the sky determination: there is a difference of  $0.7\%$ between both sky determinations. However, as the surface brightness of the cluster is much lower than the sky level ($R_{GC/Sky,z}\sim0.06$), this sky level difference results in a magnitude difference as stated above. This example again stresses the difficulty and importance of obtaining a reliable sky value. Nevertheless, the resulting magnitude errors are lower than 0.01 mag, hence the magnitude uncertainty is underestimated for this cluster. $|\Delta_{AB} ^{z^\prime}|\lesssim0.01$ for both observations.

NGC~6121 is a bright high-reddening ($E(B-V)\sim0.35$) cluster showing a $z^\prime$ magnitude difference of $\sim0.1$~mag between the short 1~s exposure obtained on May 10 2003 ($z^\prime\sim4.37$) and the 11~s exposure performed on June 6 2004 ($z^\prime\sim4.47$). This cluster is located in a very crowded field ($l\sim351\degree, b\sim16\degree$), with a vast number of stars saturating the 60~s exposures. Nevertheless, for the short exposures, only few counts are collected to determine a reliable sky value. In the case of the 1~s exposure, the sky uncertainty obtained by \textsc{mmm} was larger than the sky value itself, with $R_{GC/Sky,z}\sim1.2$ and $\Delta_{AB} ^{z^\prime}\sim-0.07$ mag, while for the 11~s exposure, $R_{GC/Sky,z}\sim1.2$ and $\Delta_{AB} ^{z^\prime}\lesssim$ 0.01 mag. On May 10 2003, the short 1~s exposure was followed by a long 410~s exposure, which could obviously not be used to determine the aperture magnitude of the cluster because of saturation issues. However, when using the long exposure to determine a more reliable sky value and then applying this sky value to the 1~s exposure, we obtain a magnitude 0.05 mag fainter than than the value obtained above. Note that this value is fully consistent with the median value of the magnitudes obtained on both May 10 2003 and June 6 2004.  

The scatter in Fig.~\ref{compdiffnight} is larger than we would expect given the known error budget (photometric, centroiding, etc). 
We add a systematic contribution of 0.03 mag for $g^{\prime}$ and 0.0435 mag for $z^{\prime}$), for the remainder of this analysis, 
to reduce the derived $\chi^2$ to 1 and account for the additional photometric uncertainty. We cannot estimate this error 
for the other bands and we therefore adopt the $z^{\prime}$ error.

\begin{table}
\centering
\caption{\label{tab:diffnight} Median magnitude differences of clusters in common on different nights. Observations on September 26, 2005 had only few observations in common with other nights.  }
\begin{tabular}{ccc}
\hline
$\Delta$ Nights & $g^\prime$& $z^\prime$ \\
\hline
5/10/03 $-$ 6/5/04 & $-0.004$ &0.015 \\
5/10/03 $-$ 6/6/04 & $-0.008$ & 0.007 \\
6/5/04 $-$ 6/6/04 & $-0.002$ & $-0.002$ \\
6/5/04 $-$ 9/26/05 &$-0.004$ & 0.001 \\
\hline 
\end{tabular} \\
\end{table}

To demonstrate this error is not caused by a systematic photometric shift of certain nights, we present in Table \ref{tab:diffnight} the median differences for clusters in common for the given nights.

\subsection{Comparing CTIO and SDSS DR9} \label{sec:CTIOSDSS}
\begin{figure}
\centering 
\includegraphics[scale=0.87,trim=2.8cm 13.1cm 9cm 6cm] {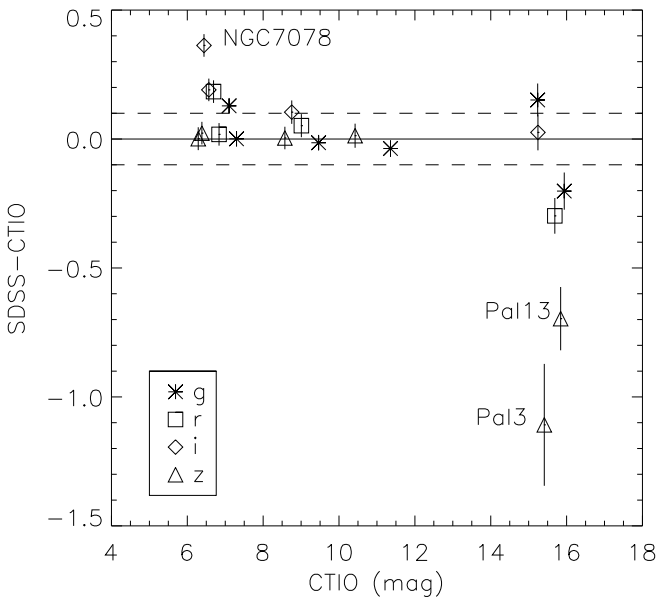}
\caption{Comparison of magnitudes based on the CTIO observations and the SDSS DR9 survey. Error bars are the combined errors on both magnitudes. }  
\label{compSDSSCTIO}
\end{figure}

The CTIO and SDSS subsamples have 6 clusters in common for the $g$- and $z$-band and 4 GCs for $r$- and $i$-band. In
Fig.~\ref{compSDSSCTIO} we compare the different magnitudes for the different filters. The included error bars are the
combined errors of both magnitudes. The RMS for the magnitude difference between CTIO and SDSS data is 0.10, 0.18, 0.21
and 0.53 mag. for the $g$, $r$, $i$ and $z$ band respectively. NGC~7078 and NGC~7089 are the outliers in the $i$-band, Pal~3 and Pal~13 are the two $z$-band outliers. These objects contribute most to the high scatter. 

Pal~3 has uncertain CTIO photometry, as was discussed in Section~\ref{sec:diffnights}. It should be noted that the CTIO $g^\prime$ magnitude, based on the observations performed on June 6 2004, is consistent with the SDSS magnitude (within the large photometric uncertainty for this faint cluster). However, $z^\prime$ magnitudes based on CTIO observations on both June 5 and 6 2004 do not compare well with the SDSS magnitude for this filter. The issues regarding the sky determination for CTIO observations of this cluster were discussed in Section~\ref{sec:diffnights}.

Pal~13 was only observed on September 26 2005 and has one of the highest specific frequencies of blue stragglers in any known GC \citep{Clark2004}. Based on the SDSS CMDs, one candidate outlier was identified. However, the star did not have proper motions, hence was not removed from the aperture photometry. This candidate outlier can not explain the $g-z$ colour difference between SDSS and CTIO: removing the star would have resulted in a $g-z$ colour correction of $-0.04$. In Section~\ref{sec:diffnights} it became clear that a small variation in the CTIO sky determination can result in a large magnitude difference, especially for faint clusters which have a lower surface brightness than the sky itself. Motivated by the latter argument, we reinspected the CTIO sky determination for Pal~13. 
The $R_{GC/Sky,g}\sim0.12$, while $R_{GC/Sky,z}\sim0.01$, so the cluster flux contribution is much smaller than the sky contribution within a half-light radius. Using \textsc{mmm} on a sky ring of 900 to 1000 pixels (referred to as method B in Section~\ref{sec:diffnights}) instead of the four corner approach (method A) results in magnitude differences of $\Delta_{AB} ^{g^\prime}\lesssim$ 0.003 mag, while these rise to $\Delta_{AB} ^{z^\prime}\sim0.35\text{ mag}$ in the $z^\prime$ band. The large photometric uncertainties are reflected in the magnitude errors, though these are smaller than the difference invoked by using the different sky estimation methods.

\begin{figure*}
\centering 
\includegraphics[scale=0.87,trim=2.8cm 13.1cm 9cm 6cm] {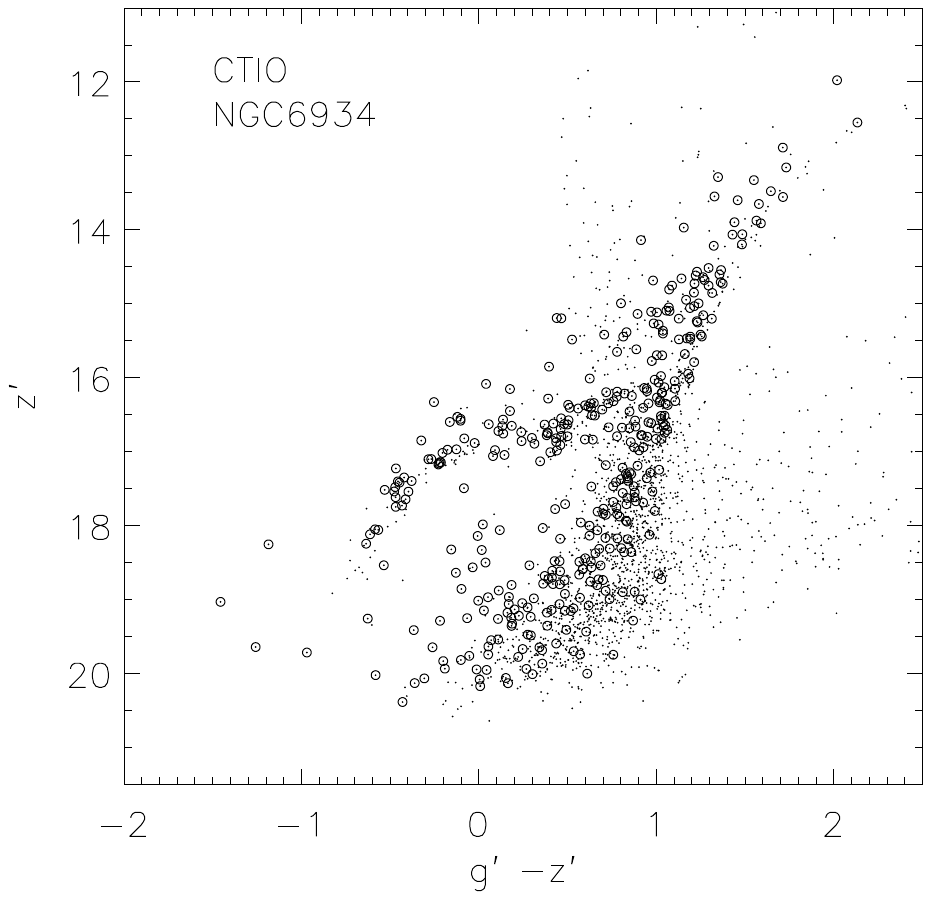}\includegraphics[scale=0.87,trim=2.8cm 13.1cm 9cm 6cm] {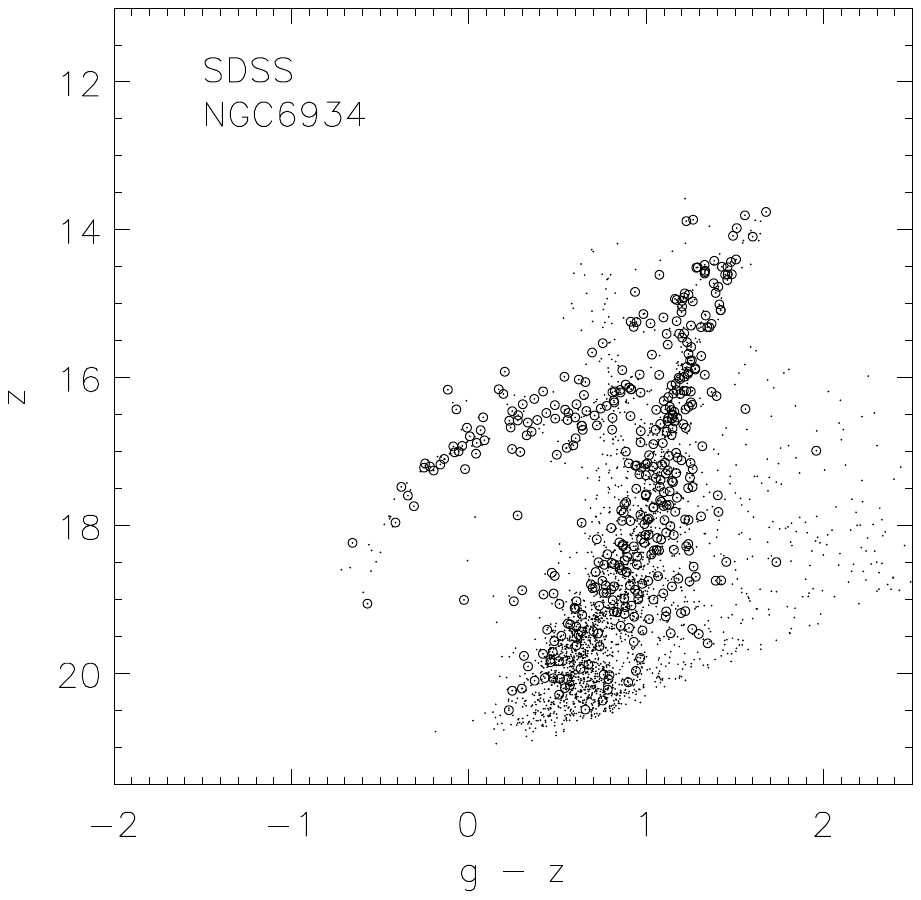}
\caption{CMDs for NGC~6934 based on CTIO and SDSS data, using open circles for stars within $r_h$, dots for stars from the entire field. It is
  clear that saturation issues in the SDSS data sweep out the tip of
  the RGB. See text for more details. }  
\label{NGC6934SDSSCTIO}
\end{figure*}

Fig. \ref{NGC6934SDSSCTIO} presents CMDs for NGC~6934 based on CTIO and SDSS data. It is clear that the RGB in the 
SDSS CMD suffers from saturation (it is known SDSS saturation starts at $r\sim14$). Moreover, more blue stragglers are 
found in the CTIO CMD and the blue HB is more extended than in the SDSS CMD. Further analysis of the CMDs will be 
presented in a subsequent paper. Nevertheless, when comparing the integrated magnitudes of NGC~6934, the agreement 
is excellent in $grz$ while in the $i$-band magnitudes show a larger difference.

The SDSS data for NGC 7078 is also saturated for the RGB stars. Only for the $z$-band the difference between SDSS and 
CTIO data is smaller than 0.01 mag, for the other filters the differences are much larger. We observed this cluster with the 
CTIO 0.9m telescope on 3 different nights in the $g^\prime$ and $z^\prime$ filters. These 3 observations deviate less 
than 0.01 mag from each other in both filters. The cluster was also observed with the $r^\prime$ and $i^\prime$ filters
but only for one night.

For the future work, we use the CTIO-based magnitudes whenever GCs have magnitudes from both subsamples, except for Pal~3 
and Pal~13, which are low-reddening clusters with very exceptional CTIO colours ($g^\prime - z^\prime\sim0$), compared to $g^\prime-z^\prime$ colours of the other low-reddening clusters ranging between 
0.4 and 1.4. We suspect issues with the sky determination for the CTIO data are causing the offsets with the SDSS data for these faint clusters.

\section{Summary} \label{sec:results}
In the current study we presented integrated photometry for 96 Galactic GCs. We discuss a variety of issues,
such as dealing with incomplete imaging (CCD cameras do not image the whole cluster), sky removal, calibration, 
the cleaning of contamination based on CMDs and proper motions and systematic errors. We obtained $g^\prime$ 
and $z^\prime$ magnitudes for about two-thirds of the Galactic GC system, making this the largest homogeneous optical
sample based on the SDSS filter system. For about half of these clusters, we also present $r^\prime$ and  $i^\prime$ photometry.

This work is the first of a series of papers, collected in the Galactic Globular Cluster Catalog (G2C2), exploiting this dataset of SDSS photometry. Future studies will deal with the colour-metallicity relations, the colour-magnitude diagrams, the spectral energy distributions, the structural
parameters and the integrated spectroscopy.

\section*{Acknowledgments}
We thankfully acknowledge the anonymous referee for very useful and thought-provoking comments. 
We would like to thank Giovanni Carraro for
fruitful discussions.  
JV acknowledges the support of ESO through a studentship. JV and MB
acknowledge the support of the Fund for Scientific Research Flanders
(FWO-Vlaanderen). AJ acknowledges support by the Chilean Ministry for
the Economy, Development, and TourismÕs Programa Iniciativa
Cient\'iÞca Milenio through grant P07-021-F, awarded to The Milky Way
Millennium Nucleus.

The authors are grateful to CTIO for the hospitality and the dedicated
assistance during the numerous observing runs. 

This research has made use of NASA's Astrophysics Data System and the
NED which is operated by the Jet Propulsion Laboratory, California
Institute of Technology, under contract with the National Aeronautics
and Space Administration. 

This research made use of Montage, funded by the National Aeronautics and Space Administration's Earth Science Technology Office, Computation Technologies Project, under Cooperative Agreement Number NCC5-626 between NASA and the California Institute of Technology. Montage is maintained by the NASA/IPAC Infrared Science Archive.

For this part of the research, we have made extensive use of the European Virtual
Observatory applications \textsc{aladin} \citep{Bonnarel2000} and \textsc{topcat}
\citep{Taylor2005}. The Virtual Observatory is a project designed to provide the 
astronomical community with the data access and the research tools necessary to
enable the exploration of the digital, multi-wavelength universe resident in the astronomical 
data archives. We used the applications provided by
astrometry.net\footnote{http://astrometry.net/}.

\bibliographystyle{mnras}
\bibliography{references}

\appendix

\section{Online tables}
\onecolumn 
\begin{center}
\begin{longtable}{lcccccccccccc}
\caption[griz]{GC $g^\prime r^\prime i^\prime z^\prime$  magnitudes and errors.  } \label{appendixtab:griz} \\
\hline \multicolumn{1}{|c}{ID} &
\multicolumn{1}{c}{$g^\prime$} &
\multicolumn{1}{c}{$\sigma_{g^\prime}$ } &
\multicolumn{1}{c}{$r^\prime$} &
\multicolumn{1}{c}{$\sigma_{r^\prime}$}&
\multicolumn{1}{c}{$i^\prime$}&
\multicolumn{1}{c}{$\sigma_{i^\prime}$}&
\multicolumn{1}{c}{$z^\prime$}&
\multicolumn{1}{c}{$\sigma_{z^\prime}$}&
\multicolumn{1}{c}{CMD$_{g^\prime}$}&
\multicolumn{1}{c}{CMD$_{r^\prime}$}&
\multicolumn{1}{c}{CMD$_{i^\prime}$}&
\multicolumn{1}{c|}{CMD$_{z^\prime}$}\\ \hline 
\endfirsthead

\multicolumn{5}{c}
{{\bfseries \tablename\ \thetable{} -- continued from previous page}} \\
\hline \multicolumn{1}{|c}{ID} &
\multicolumn{1}{c}{$g^\prime$} &
\multicolumn{1}{c}{$\sigma_{g^\prime}$ } &
\multicolumn{1}{c}{$r^\prime$} &
\multicolumn{1}{c}{$\sigma_{r^\prime}$}&
\multicolumn{1}{c}{$i^\prime$}&
\multicolumn{1}{c}{$\sigma_{i^\prime}$}&
\multicolumn{1}{c}{$z^\prime$}&
\multicolumn{1}{c}{$\sigma_{z^\prime}$}&
\multicolumn{1}{c}{CMD$_{g^\prime}$}&
\multicolumn{1}{c}{CMD$_{r^\prime}$}&
\multicolumn{1}{c}{CMD$_{i^\prime}$}&
\multicolumn{1}{c|}{CMD$_{z^\prime}$}\\ \hline 
\endhead

\hline \multicolumn{13}{|r|}{{Continued on next page}} \\ \hline
\endfoot

\hline \hline
\endlastfoot
 NGC104     &      4.912     &      0.030     & ...     & ...     & ...     & ...     &      3.677     &      0.044     &      0.00     & ...     & ...     &      0.00 \\
 NGC288     &      9.080     &      0.032     &      8.600     &      0.045     &      8.295     &      0.044     &      8.139     &      0.044     &      0.00     &      0.00     &      0.00     &      0.00 \\
 NGC362     &      7.471     &      0.030     &      6.925     &      0.043     &      6.618     &      0.043     &      6.419     &      0.043     &      0.01     &      0.02     &      0.02     &      0.03 \\
NGC1261     &      9.474     &      0.031     &      8.995     &      0.045     &      8.744     &      0.045     &      8.582     &      0.045     &      0.03     &      0.05     &      0.06     &      0.06 \\
    AM1     &     15.958     &      0.035     &     15.582     &      0.046     &     15.219     &      0.047     &     15.159     &      0.056     &      0.07     &      0.13     &      0.14     &      0.16 \\
NGC1851     &      8.280     &      0.031     &      7.703     &      0.044     &      7.413     &      0.044     &      7.194     &      0.044     &      0.03     &      0.04     &      0.05     &      0.05 \\
NGC1904     &      9.006     &      0.030     &      8.570     &      0.044     &      8.351     &      0.044     &      8.178     &      0.044     &      0.02     &      0.04     &      0.05     &      0.05 \\
NGC2298     &      9.650     &      0.039     &      9.287     &      0.050     &      9.036     &      0.052     &      8.865     &      0.054     &      0.06     &      0.10     &      0.10     &      0.10 \\
NGC2808     &      6.695     &      0.030     &      6.194     &      0.044     &      5.916     &      0.044     &      5.725     &      0.044     &      0.00     &      0.00     &      0.00     &      0.00 \\
     E3     &     10.787     &      0.041     & ...     & ...     & ...     & ...     &     10.368     &      0.072     &      0.00     & ...     & ...     &      0.00 \\
   Pal3     &     15.238     &      0.044     & ...     & ...     & ...     & ...     &     15.412     &      0.232     &      0.00     & ...     & ...     &      0.00 \\
NGC3201     &      7.006     &      0.033     & ...     & ...     & ...     & ...     & ...     & ...     &      0.01     & ...     & ...     &      0.00 \\
NGC4372     &      6.489     &      0.031     & ...     & ...     & ...     & ...     &      5.796     &      0.044     &      0.00     & ...     & ...     &      0.00 \\
 Rup106     &     11.981     &      0.031     & ...     & ...     & ...     & ...     &     11.204     &      0.168     &      0.59     & ...     & ...     &      0.69 \\
NGC4590     &      8.903     &      0.037     & ...     & ...     & ...     & ...     &      8.127     &      0.054     &      0.00     & ...     & ...     &      0.00 \\
NGC4833     &      7.034     &      0.034     & ...     & ...     & ...     & ...     & ...     & ...     &      0.03     & ...     & ...     &      0.02 \\
NGC5139     &      4.131     &      0.030     & ...     & ...     & ...     & ...     &      3.198     &      0.044     &      0.01     & ...     & ...     &      0.01 \\
NGC5286     &      7.551     &      0.030     & ...     & ...     & ...     & ...     &      6.664     &      0.044     &      0.00     & ...     & ...     &      0.00 \\
NGC5634     &     10.060     &      0.030     & ...     & ...     & ...     & ...     &      9.205     &      0.046     &      0.00     & ...     & ...     &      0.00 \\
NGC5694     &     10.781     &      0.030     & ...     & ...     & ...     & ...     &      9.899     &      0.044     &      0.00     & ...     & ...     &      0.00 \\
 IC4499     &     10.256     &      0.032     &      9.820     &      0.044     &      9.520     &      0.044     &      9.372     &      0.044     &      0.18     &      0.24     &      0.23     &      0.22 \\
NGC5824     &      9.321     &      0.030     &      8.876     &      0.043     &      8.648     &      0.043     &      8.477     &      0.043     &      0.00     &      0.00     &      0.00     &      0.00 \\
NGC5897     &      9.171     &      0.031     &      8.835     &      0.044     &      8.542     &      0.044     &      8.414     &      0.044     &      0.03     &      0.04     &      0.04     &      0.04 \\
NGC5904     &      6.861     &      0.030     & ...     & ...     & ...     & ...     &      5.867     &      0.045     &      0.00     & ...     & ...     &      0.00 \\
NGC5927     &      7.785     &      0.030     &      7.184     &      0.043     &      6.816     &      0.044     &      6.521     &      0.045     &      0.00     &      0.00     &      0.00     &      0.00 \\
NGC5946     &      8.340     &      0.031     &      7.889     &      0.044     &      7.645     &      0.045     &      7.477     &      0.045     &      0.30     &      0.14     &      0.08     &      0.05 \\
NGC5986     &      7.794     &      0.031     &      7.349     &      0.045     &      7.072     &      0.044     &      6.894     &      0.044     &      0.06     &      0.03     &      0.02     &      0.02 \\
NGC6093     &      7.924     &      0.030     &      7.417     &      0.044     &      7.148     &      0.044     &      6.939     &      0.044     &      0.00     &      0.00     &      0.00     &      0.00 \\
NGC6121     &      5.251     &      0.031     & ...     & ...     & ...     & ...     &      4.435     &      0.044     &      0.03     & ...     & ...     &      0.02 \\
NGC6101     &     10.171     &      0.032     &      9.736     &      0.045     &      9.460     &      0.044     &      9.301     &      0.046     &      0.00     &      0.00     &      0.00     &      0.00 \\
NGC6144     &      7.876     &      0.031     &      7.811     &      0.044     &      7.689     &      0.044     &      7.602     &      0.044     &      0.19     &      0.13     &      0.09     &      0.07 \\
NGC6139     &      7.375     &      0.030     & ...     & ...     & ...     & ...     &      6.623     &      0.044     &      0.00     & ...     & ...     &      0.00 \\
NGC6171     &      8.084     &      0.032     & ...     & ...     & ...     & ...     &      7.109     &      0.050     &      0.00     & ...     & ...     &      0.00 \\
NGC6218     &      7.691     &      0.036     & ...     & ...     & ...     & ...     &      6.653     &      0.053     &      0.00     & ...     & ...     &      0.00 \\
NGC6235     &      9.542     &      0.036     & ...     & ...     & ...     & ...     &      8.552     &      0.064     &      0.00     & ...     & ...     &      0.00 \\
NGC6254     &      6.842     &      0.031     & ...     & ...     & ...     & ...     & ...     & ...     & ...     & ...     & ...     & ... \\
NGC6273     &      6.909     &      0.031     & ...     & ...     & ...     & ...     &      5.801     &      0.044     &      0.00     & ...     & ...     &      0.00 \\
NGC6284     &      9.247     &      0.030     & ...     & ...     & ...     & ...     &      8.145     &      0.044     &      0.00     & ...     & ...     &      0.00 \\
NGC6287     &      8.121     &      0.030     &      7.811     &      0.046     &      7.580     &      0.045     &      7.426     &      0.044     &      0.00     &      0.00     &      0.00     &      0.00 \\
NGC6293     &      7.440     &      0.030     & ...     & ...     & ...     & ...     &      6.999     &      0.044     &      0.00     & ...     & ...     &      0.05 \\
NGC6304     &      7.677     &      0.030     & ...     & ...     & ...     & ...     &      6.313     &      0.044     &      0.00     & ...     & ...     &      0.00 \\
NGC6316     &      8.131     &      0.030     & ...     & ...     & ...     & ...     &      7.090     &      0.043     &      0.00     & ...     & ...     &      0.00 \\
NGC6333     &      7.467     &      0.032     & ...     & ...     & ...     & ...     &      6.676     &      0.044     &      0.00     & ...     & ...     &      0.00 \\
NGC6342     &      9.327     &      0.031     & ...     & ...     & ...     & ...     &      8.231     &      0.044     &      0.00     & ...     & ...     &      0.00 \\
NGC6356     &      8.499     &      0.031     & ...     & ...     & ...     & ...     &      7.231     &      0.048     &      0.01     & ...     & ...     &      0.03 \\
NGC6355     &      7.413     &      0.039     & ...     & ...     & ...     & ...     &      6.908     &      0.049     &      0.00     & ...     & ...     &      0.00 \\
NGC6352     &      7.788     &      0.035     & ...     & ...     & ...     & ...     &      6.562     &      0.044     &      0.00     & ...     & ...     &      0.00 \\
 IC1257     & ...     & ...     &     10.927     &      0.045     &     10.719     &      0.046     & ...     & ...     &      ...    &      0.12     &      0.13     &      ...\\
NGC6366     &      7.736     &      0.031     & ...     & ...     & ...     & ...     &      6.769     &      0.044     &      0.23     & ...     & ...     &      0.10 \\
NGC6362     &      8.429     &      0.031     & ...     & ...     & ...     & ...     &      7.334     &      0.044     &      0.00     & ...     & ...     &      0.00 \\
NGC6388     &      6.962     &      0.030     & ...     & ...     & ...     & ...     &      5.754     &      0.044     &      0.00     & ...     & ...     &      0.00 \\
NGC6402     &      7.503     &      0.030     & ...     & ...     & ...     & ...     &      6.203     &      0.044     &      0.00     & ...     & ...     &      0.00 \\
NGC6401     &      7.373     &      0.036     & ...     & ...     & ...     & ...     & ...     & ...     & ...     & ...     & ...     & ... \\
NGC6397     & ...     & ...     & ...     & ...     & ...     & ...     &      5.450     &      0.044     &      0.00     & ...     & ...     &      0.00 \\
NGC6426     &     11.051     &      0.031     & ...     & ...     & ...     & ...     &     10.030     &      0.044     &      0.00     & ...     & ...     &      0.00 \\
NGC6440     &      7.504     &      0.030     & ...     & ...     & ...     & ...     &      6.080     &      0.044     &      0.00     & ...     & ...     &      0.00 \\
NGC6441     &      6.635     &      0.030     & ...     & ...     & ...     & ...     &      5.700     &      0.044     &      0.00     & ...     & ...     &      0.00 \\
NGC6453     &      9.016     &      0.031     &      8.381     &      0.044     & ...     & ...     & ...     & ...     & ...     & ...     & ...     & ... \\
NGC6496     &     10.095     &      0.055     & ...     & ...     & ...     & ...     & ...     & ...     &      0.49     & ...     & ...     &      0.55 \\
NGC6517     &      8.298     &      0.030     &      7.918     &      0.043     &      7.613     &      0.044     &      7.411     &      0.044     &      0.00     &      0.00     &      0.00     &      0.00 \\
NGC6539     &      7.740     &      0.036     &      7.148     &      0.044     &      6.738     &      0.044     &      6.457     &      0.045     &      0.22     &      0.14     &      0.12     &      0.11 \\
NGC6544     &      3.941     &      0.033     &      4.341     &      0.045     & ...     & ...     & ...     & ...     & ...     & ...     & ...     & ... \\
NGC6541     &      7.262     &      0.031     & ...     & ...     & ...     & ...     & ...     & ...     &      0.00     & ...     & ...     &      0.01 \\
NGC6553     &      5.254     &      0.030     &      5.145     &      0.044     &      4.966     &      0.044     &      4.822     &      0.044     &      0.19     &      0.09     &      0.06     &      0.04 \\
NGC6558     &      7.751     &      0.035     &      7.238     &      0.045     &      6.976     &      0.045     &      6.866     &      0.044     &      0.00     &      0.00     &      0.00     &      0.00 \\
 IC1276     &      6.814     &      0.037     &      6.427     &      0.044     &      6.148     &      0.044     &      6.016     &      0.045     &      0.19     &      0.14     &      0.18     &      0.22 \\
NGC6569     &      8.801     &      0.031     &      8.048     &      0.044     &      7.605     &      0.044     &      7.278     &      0.044     &      0.05     &      0.04     &      0.03     &      0.04 \\
NGC6584     &      9.793     &      0.032     &      9.279     &      0.051     &      9.017     &      0.048     &      8.947     &      0.045     &      0.00     &      0.00     &      0.00     &      0.00 \\
NGC6624     &      8.341     &      0.031     &      7.696     &      0.044     &      7.285     &      0.046     &      7.009     &      0.049     &      0.00     &      0.00     &      0.00     &      0.00 \\
NGC6626     &      6.180     &      0.030     &      5.706     &      0.044     &      5.424     &      0.044     & ...     & ...     & ...     & ...     & ...     & ... \\
NGC6638     &      8.977     &      0.031     &      8.351     &      0.044     &      8.033     &      0.044     &      7.849     &      0.044     &      0.00     &      0.00     &      0.00     &      0.00 \\
NGC6637     &      8.295     &      0.064     &      7.693     &      0.048     &      7.306     &      0.049     & ...     & ...     &      0.00     &      0.00     &      0.00     &      0.00 \\
NGC6642     &      9.309     &      0.032     & ...     & ...     & ...     & ...     &      8.152     &      0.046     &      0.11     & ...     & ...     &      0.12 \\
NGC6652     &      9.818     &      0.031     & ...     & ...     & ...     & ...     &      8.737     &      0.046     &      0.00     & ...     & ...     &      0.00 \\
   Pal8     &     10.569     &      0.031     & ...     & ...     & ...     & ...     &      9.349     &      0.045     &      0.01     & ...     & ...     &      0.11 \\
NGC6681     &      8.859     &      0.030     & ...     & ...     & ...     & ...     & ...     & ...     &      0.00     & ...     & ...     &      0.00 \\
NGC6712     &      7.955     &      0.030     & ...     & ...     & ...     & ...     & ...     & ...     &      0.00     & ...     & ...     &      0.00 \\
NGC6715     &      7.973     &      0.030     & ...     & ...     & ...     & ...     &      6.946     &      0.043     &      0.00     & ...     & ...     &      0.00 \\
NGC6723     &      7.585     &      0.031     & ...     & ...     & ...     & ...     &      6.810     &      0.045     &      0.00     & ...     & ...     &      0.00 \\
NGC6749     &      6.735     &      0.045     & ...     & ...     & ...     & ...     &      6.341     &      0.044     &      0.41     & ...     & ...     &      0.05 \\
NGC6760     &      8.304     &      0.030     & ...     & ...     & ...     & ...     &      6.522     &      0.044     &      0.00     & ...     & ...     &      0.00 \\
NGC6779     &      8.613     &      0.030     &      8.138     &      0.044     &      7.879     &      0.044     &      7.748     &      0.044     &      0.00     &      0.00     &      0.00     &      0.00 \\
Terzan7     &     12.469     &      0.031     & ...     & ...     & ...     & ...     &     11.567     &      0.044     & 0.00    & ...     & ...     & 0.00 \\
  Pal10     &      7.862     &      0.039     &      7.498     &      0.050     &      7.328     &      0.056     &      7.216     &      0.047     &      1.33     &      0.62     &      0.39     &      0.33 \\
  Pal11     &     11.439     &      0.034     & ...     & ...     & ...     & ...     &     10.413     &      0.048     &      0.00     & ...     & ...     &      0.00 \\
NGC6838     &      7.973     &      0.039     & ...     & ...     & ...     & ...     & ...     & ...     &      0.00     & ...     & ...     &      0.00 \\
NGC6864     &      9.167     &      0.030     &      8.608     &      0.044     &      8.320     &      0.044     &      8.099     &      0.044     &      0.00     &      0.00     &      0.00     &      0.00 \\
NGC6934     &      9.460     &      0.030     &      9.007     &      0.045     &      8.754     &      0.045     &      8.572     &      0.044     &      0.00     &      0.00     &      0.00     &      0.00 \\
NGC6981     &     10.132     &      0.031     &      9.666     &      0.045     &      9.391     &      0.045     &      9.213     &      0.046     &      0.00     &      0.00     &      0.00     &      0.00 \\
NGC7006     &     11.357     &      0.030     & ...     & ...     & ...     & ...     &     10.423     &      0.045     &      0.03     & ...     & ...     &      0.02 \\
NGC7078     &      7.101     &      0.030     &      6.692     &      0.044     &      6.439     &      0.044     &      6.288     &      0.044     &      0.00     &      0.00     &      0.00     &      0.00 \\
NGC7089     &      7.297     &      0.030     &      6.836     &      0.044     &      6.567     &      0.044     &      6.384     &      0.044     &      0.00     &      0.01     &      0.00     &      0.00 \\
NGC7099     &      8.306     &      0.030     & ...     & ...     & ...     & ...     &      7.501     &      0.044     &      0.00     & ...     & ...     &      0.00 \\
  Pal12     &     12.489     &      0.030     &     11.826     &      0.043     &     11.482     &      0.044     &     11.345     &      0.045     &      1.38     &      1.24     &      1.15     &      1.06 \\
  Pal13     &     15.938     &      0.040     &     15.690     &      0.048     &     15.244     &      0.051     &     15.843     &      0.112     &      0.00     &      0.00     &      0.00     &      0.00 \\
NGC7492     &     12.197     &      0.038     &     11.824     &      0.045     &     11.570     &      0.044     &     11.560     &      0.046     &      0.00     &      0.00     &      0.00     &      0.00 \\

\end{longtable}
\end{center}

\label{lastpage}
\end{document}